\documentclass[prd,aps,twocolumn,amsmath,amssymb,nofootinbib,preprintnumbers]
{revtex4}

\usepackage{graphicx}% Include figure files
\usepackage{dcolumn}% Align table columns on decimal point
\usepackage{bm}% bold math
\usepackage{amsmath}
\usepackage{amsfonts}
\usepackage{euscript,bbm}
\usepackage{ifthen}
\usepackage{psfrag}
\usepackage{slashed}

\def\ls{\mathrel{\lower4pt\vbox{\lineskip=0pt\baselineskip=0pt
           \hbox{$<$}\hbox{$\sim$}}}}
\def\gs{\mathrel{\lower4pt\vbox{\lineskip=0pt\baselineskip=0pt
           \hbox{$>$}\hbox{$\sim$}}}}
\def\drawbox#1#2{\hrule height#2pt

\hbox{\vrule width#2pt height#1pt \kern#1pt
              \vrule width#2pt}
              \hrule height#2pt}

\def\Asym#1#2{\vcenter{\vbox{\drawbox{#1}{#2}
              \kern-#2pt       % line up boxes
              \drawbox{#1}{#2}}}}

\newcommand{\Expect}[1]{\left\langle #1 \right\rangle}
\newcommand{\be}{\begin{equation}}
\newcommand{\ee}{\end{equation}}
\newcommand{\bea}{\begin{eqnarray}}
\newcommand{\eea}{\end{eqnarray}}

\begin{document}

%
%\vspace*{2cm}
\title{Non-thermal Higgsino Dark  Matter: Cosmological Motivations and Implications for a $125$ GeV Higgs}

\author{Rouzbeh Allahverdi$^{1}$}
\author{Bhaskar Dutta$^{2}$}
\author{Kuver Sinha$^{2}$}

\affiliation{$^{1}$~Department of Physics and Astronomy, University of New Mexico, Albuquerque, NM 87131, USA \\
$^{2}$~ Mitchell Institute for Fundamental Physics and Astronomy\\
Department of Physics, Texas A\&M University, College Station, TX 77843-4242, USA}

\begin{abstract}

If the lightest supersymmetric particle (LSP) is Higgsino-like, the thermal relic density is lower than the observed dark matter content for a LSP mass in the sub-TeV region. We outline constraints arising from the Fermi Gamma-Ray Telescope data and LSP production from gravitino decay that must be satisfied by a successful non-thermal Higgsino scenario. We show that in a generic class of models where anomaly and modulus mediated contributions to supersymmetry breaking are of comparable size, Higgsino arises as the only viable sub-TeV dark matter candidate if gravitinos are heavy enough to decay before the onset of big bang nucleosynthesis (BBN). The correct relic density can be obtained via modulus decay in these models. As an explicit example, we consider a modulus sector in effective field theory ($D=4,~N=1$ supergravitiy arising from type IIB KKLT compactification). Within this class of mirage mediation models, heaviness of the gravitino forces a sub-TeV Higgsino LSP and gives a Higgs mass around $125$ GeV. In this example, the constraints from direct detection experiments are also satisfied.
\end{abstract}
MIFPA-12-28,
CETUP$^\ast$-12/007\\

\maketitle

%%%%%%%%%%%%%%%%%%%%%%%%

\section{Introduction}

Supersymmetry not only stabilizes the Higgs mass against quantum corrections, it also provides a candidate for dark matter. In $R$-parity conserving models the lightest supersymmetric particle (LSP) is stable, hence a dark matter candidate. The lightest neutralino, which is a mixture of Bino, Wino and Higgsinos, is the most suitable dark matter candidate with the prospect for detection in various direct and indirect searches.

In this work we point out that a comprehensive solution to the cosmological gravitino problem motivates the dark matter to be Higgsino-like. Gravitinos heavier than ${\cal O}(40)$ TeV have a lifetime shorter than 0.1 s and decay before the onset of big bang nucleosynthesis (BBN). This results in a considerable relaxation as the gravitino abundance will not be subject to tight BBN bounds~\cite{BBN}.

%In modulus mediation models, the masses of the Bino and Wino are sensitive to the mass of the gravitino $m_{3/2}$  \cite{Conlon:2006us, }.
In effective supergravity, the masses of the Bino and Wino are sensitive to the mass of the gravitino $m_{3/2}$  \cite{Kaplunovsky:1993rd}, and in particular, for $m_{3/2} > 40$ TeV, one typically has Bino and Wino masses above TeV in type IIB modulus mediation models. On the other hand, the Higgsino mass depends on the $\mu$ parameter, which can be reduced by anomaly mediated contribution to supersymmetry breaking.
%which is required to be more than $\mathcal{O}(50)$ TeV  for the gravitinos to decay well before  Big Bang Nucleosynthesis.
As a result, if we demand that the dark matter particle has a mass in the sub-TeV region, the Higgsino becomes a more natural candidate.
%since its mass can be reduced by anomaly mediated contributions to supersymmetry breaking.

If the lightest neutralino is pre-dominantly Higgsino, with mass in the sub-TeV region, the annihilation rate is typically larger than the nominal value $\langle \sigma_{\rm ann} v \rangle = 3 \times 10^{-26}$ ${\rm cm^3 s^{-1}}$, thus resulting in an insufficient thermal relic abundance~\cite{Baer:2012uy}. A natural way to obtain the correct dark matter relic density is to consider non-thermal sources of Higgsino production.

We consider scenarios where Higgsino dark matter is non-thermally produced by a late decaying modulus \cite{Moroi:1999zb}. We find that for annihilation rate to be compatible with bounds from the Fermi Gamma-Ray Telescope~\cite{fermi}, the modulus decay should reheat the universe to a temperature $T_{\rm d} \sim {\cal O}({\rm GeV})$.
%The decay of moduli fields stabilized via non-perturbative schemes can lead to a successful and consistent scenario of Higgsino production in this framework.
An additional requirement is that the branching ratio for modulus decay to the gravitino is $\ls {\cal O}(10^{-5})$, so that the decay of gravitinos thus produced does not lead to dark matter overproduction.

As an example of the modulus sector, we consider the standard scenario of KKLT compactification \cite{Kachru:2003aw}, with the Kahler modulus reheating the universe around $1$ GeV. Within this framework, for appropriate values of the relative contributions of anomaly and modulus mediated contributions, Higgsino emerges as the dark matter candidate. The annihilation rate is consistent with the Fermi bounds, and the correct relic density is obtained
%by the enhancement of the thermal relic density due to
by non-thermal production. The Higgs mass $m_h \sim 125$ GeV \footnote{Recent experiments at the LHC have provided strong hints of a Higgs-like particle at $\sim 125$ GeV \cite{LHCHiggs}.} is also satisfied in this scenario, and we find that it actually requires the gravitino mass to be in the cosmologically safe region.  Moreover, the spin independent scattering cross section is consistent with the latest bounds from direct detection experiments~\cite{direct}.

Within this specific example, however, decay of the gravitinos that are directly produced from modulus decay overproduces dark matter. This is a direct consequence of the couplings between the modulus and the helicity $\pm 1/2$ components of the gravitino, which are in turn set by the underlying K\"ahler geometry of the effective $D=4,~N=1$ supergravity theory.
We summarize a set of geometric conditions in the modulus sector that are sufficient to ensure consistent non-thermal Higgsino dark matter as outlined above.

We note that apart from the purely cosmological motivations shown in this study, the Higgsino also emerges as the LSP within the framework of Natural Supersymmetry as discussed in \cite{Papucci:2011wy}, \cite{Hall:2011aa}, \cite{Baer:2012uy}.

The paper is organized as follows. In Section \ref{Gravproblem}, we relate the cosmological gravitino problem with the preference for Higgsino dark matter. In Section \ref{nonthermal}, we outline the conditions that must be satisfied by any successful scenario of non-thermal Higgsino dark matter. In Section IV, we work out an explicit example of a non-thermal scenario. In Section \ref{gravitino}, we outline the general constraints on an effective modulus sector in order to avoid overproduction of gravitinos. We conclude the paper in Section VI.

\section{Cosmological Gravitino Problem and Higgsino Dark Matter} \label{Gravproblem}

In this section, we discuss the cosmological gravitino problem, and argue that requiring dark matter in the sub-TeV range makes Higgsino a natural candidate due to an interplay between modulus and anomaly mediation contributions.

\subsection{Cosmological Gravitino Constraint}

The decay width of a particle $\phi$, which may be a modulus or the gravitino, with couplings of gravitational strength to the visible sector fields is
\be \label{decaywidth}
\Gamma_{\phi} = \frac{c}{2\pi} \frac{m_\phi^3}{M^2_{\rm P}} ,
\ee
where $c$ depends on the couplings of the decaying field. For moduli fields, we typically have $c \sim 0.1-1$. For gravitinos, $c$ can be computed explicitly since supersymmetry fixes the couplings of the gravitino to the visible sector. One has a maximal value of $c \sim 1.5$ in this case \cite{Moroi:1995fs}.

The decay occurs when $H \simeq \Gamma_{\phi}$, with $H$ being the Hubble expansion rate of the universe. Modulus decay reheats the universe to the following temperature
\begin{eqnarray} \label{Td}
T_{\rm d} \simeq
%\left( \frac{10.75 }{g_*}\right)^{1/4} \sqrt{\Gamma_\tau M_p} \nonumber \\
(5 ~ {\rm MeV}) ~ c^{1/2} \left(\frac{10.75}{g_*}\right)^{1/4} \left(\frac{m_\phi}{100~{\rm TeV}}\right)^{3/2} \, , \nonumber \\
\end{eqnarray}
where $g_*$ is the total number of relativistic degrees of freedom at $T_{\rm d}$ ($g_* = 10.75$ for $T_{\rm d} \gtrsim {\cal O}({\rm MeV})$).
%This results in:
%
%\bea \label{dilution}
%{3 T_{\rm d} \over 4 m_\phi} \sim
%\,\, \frac{3}{4} \frac{T_{\rm r}}{m_\tau} \nonumber \\
%5 \times 10^{-8} ~ \left(\frac{m_\phi}{100\, {\rm TeV}}\right)^{1/2} \, .
%\eea
%

%For the case of gravitinos, one typically has a maximal value of $c_{3/2} \sim 1.5$ \cite{Moroi:1995fs}, which may be computed explicitly since %supersymmetry fixes the couplings of the gravitino to the supercurrent.
Gravitinos that have a lifetime shorter than 0.1 s decay before the onset of BBN and avoid any conflict with its successful predictions. Such alifetime corresponds to $T_{\rm d} \gs 3$ MeV, which requires that $m_{3/2} \gs {\cal O}(40)$ TeV from Eq.~(\ref{Td}).
%One thus requires gravitinos with $m_{3/2} \sim \mathcal{O}(50)$ TeV or more to avoid gravitino decay from conflicting with the successes of Big Bang %Nucleosynthesis, which occurs around $2-5$ MeV
%\cite{Banks:1995dt}.

\subsection{Higgsino LSP}

To obtain a TeV scale spectrum in the observable sector for such heavy gravitinos, one needs to consider models where there is a hierarchy between the gravitino and the other superpartner masses. We consider modulus mediation, where supersymmetry is broken by a gravitationally coupled modulus in the hidden sector. An example of a broad class of models that provide the required hierarchy is in compactifications of Type IIB string theory, with the following ingredients: $(a)$ a K\"ahler modulus $T_a$ stabilized by non-perturbative effects $(b)$ complex structure moduli stabilized by fluxes $(c)$ visible sector on $D7$ branes \cite{Conlon:2006us}. In such cases, the gaugino masses obey
\be \label{hierarchy}
M_{\tilde g} \, \approx \, \frac{m_{3/2}}{\ln{(M_{\rm P}/m_{3/2})}} \,\,.
\ee

When anomaly mediated contributions are subdominant, the Bino is the LSP. However, for $m_{3/2} > 40$ TeV, the Bino mass is typically above $\mathcal{O}(1)$ TeV as seen from Eq.~(\ref{hierarchy}). We will demonstrate this in our explicit example later.

%
%Typically, one obtains GUT scale gaugino masses of $\mathcal{O}(1)$ TeV, for $m_{3/2} \sim 50$ TeV.
%This estimate may be different for other classes of models, but the important thing to note is that the gravitino mass is usually most directly connected to the gauginos, and a heavier gravitino pulls up the gaugino spectrum.
%
For the Higgsinos, there is an additional freedom. Starting with equal modulus mediated contributions, anomaly mediation lowers the mass of the gluino, while increasing the mass of the Bino and Wino.
%Denoting the relative strength of anomaly and modulus mediated contributions by $\alpha\,\equiv\,\frac{m_{3/2}}{M_0\ln(M_{Pl}/m_{3/2})}$, where $M_0$ %is the modulus mediated contribution at the GUT scale,
In this case, one has
%
%affect the low-energy mass of the gluino differently from the way it affects the Wino or Bino.
%
%Anticipating our example, if we denote the relative strength of anomaly mediation by $\alpha$, with $\alpha \rightarrow 0$ being the limit where anomaly contributions become negligible, one has roughly
%
\bea
&M_3 &: M_2 : M_1  \nonumber \\
&\sim & (1 - 0.3 \alpha)g_3^2 : (1+ 0.1 \alpha)g_2^2 : (1+ 0.66 \alpha)g_1^2 \,\, ,
\eea
where $M_0$ is the modulus mediated contribution at the GUT scale, $\alpha \equiv m_{3/2}/M_0\ln(M_{\rm P}/m_{3/2})$ denotes the relative strength of anomaly and modulus mediated contributions, and $g_{1,2,3}$ are the gauge coupling constants.
%The functions $f(b_i)$ are proportional to the beta function coefficients, and depend on how one parametrizes $\alpha$.
\cite{Choi:2006im, Choi:2005uz}. The limit $\alpha \rightarrow 0$ corresponds to vanishing anomaly mediated contribution and Bino-like LSP.

The value of the Higgsino mass parameter $\mu$ depends on the low-scale value of $m^2_{H_u}$, which is mainly driven by the gluino mass \cite{Endo:2005uy}. Increasing $\alpha$ lowers the gluino mass, which in turn lowers the low-scale value of $m^2_{H_u}$ due to the top Yukawa coupling.
%between $H_u$ and the stop.
On the other hand, the Bino and the Wino become heavier.

Thus, if we demand the dark matter candidate mass to be less than $\mathcal{O}(1)$ TeV, Higgsino LSP is preferred by the cosmological gravitino problem, provided that the anomaly mediated contribution to the soft masses competes with the modulus mediated contribution.
%This is most clearly realized in mirage mediation models with dilaton-modulus mixing in the $D7$ gauge kinetic function, which can be obtained by non-zero gauge flux on the $D7$.
%

The mass hierarchy between the scalar masses and the gravitino mass is more model dependent, and depends on the curvature properties of the underlying K\"ahler manifold. Stop masses of $\mathcal{O}(1)$ TeV, preferred by a $125$ GeV Higgs mass in the MSSM, are obtained in models with heavy gravitinos where the suppression is similar to Eq.~(\ref{hierarchy}).

%In models where the scalar spectrum suffers suppression with respect to the gravitino at a level similar to that of the gauginos,

%the cosmological gravitino problem drives the scalars to be in the $\mathcal{O}(1-2)$ TeV range. A Higgs mass of $125$ GeV supports this in the context of the MSSM, since it requires a lower bound of $\sim \mathcal{O}(1)$ TeV for stop masses.

%The scalar masses depend on the holomorphic bisectional curvature of the plane (in tangent space) spanned by the scalars and the supersymmetry breaking modulus.

\section{Scenarios of Non-thermal Higgsinos Production} \label{nonthermal}

In this section, we consider non-thermal production of Higgsino dark matter. Standard relic density calculations predict thermal underproduction of Higgsinos for masses less than about a TeV \cite{Baer:2012uy}. Thus, one is motivated to study non-thermal scenarios where enhancement of the relic density occurs naturally. Non-thermal production is inevitable if the modulus $\phi$ that participates in supersymmetry breaking, or any other modulus, reheats the universe at a scale below the dark matter freeze-out temperature $T_{\rm f} \sim m_\chi/25$.

The dark matter relic density from modulus decay is given by
\be \label{dmdensity}
{n_\chi \over s} \approx 5 \times 10^{-10} ~ \left({1 ~ {\rm GeV} \over m_\chi} \right) ~ {3 \times 10^{-26} ~ {\rm cm^3 ~ s^{-1}} \over \langle \sigma_{\rm ann} v \rangle_{\rm f}} ~ \left({T_{\rm f} \over T_{\rm d}}\right) ,
\ee
where
%$T_{\rm f} \sim m_\chi/20$ is the freeze-out temperature for dark matter annihilation and
$\langle \sigma_{\rm ann} v \rangle_{\rm f}$ is the annihilation rate at the time of freeze-out.

The Higssino DM mainly annihilates into heavy Higgs bosons, $W$ bosons and $t$ quarks via $S$-wave annihilation if $m_\chi$ has necessary phase space for these particles to be produced. The $S$-wave nature of the annihilation implies that annihilation rate at the freeze-out time is essentially the same as that at the present time. The latter is constrained by the gamma ray flux from dwarf spheriodal galaxies~\cite{fermi}:
\begin{eqnarray} \label{fermi}
&& \langle \sigma_{\rm ann} v \rangle_f \ls 10^{-25} ~ {\rm cm}^3 ~ {\rm s}^{-1} ~ ~ ~ ~ ~ ~ ~ ~ ~ ~ m_\chi = 100 ~ {\rm GeV} \, , \nonumber \\
&& \langle \sigma_{\rm ann} v \rangle_f \ls 3 \times 10^{-24} ~ {\rm cm}^3 ~ {\rm s}^{-1} ~ ~ ~ ~ ~ m_\chi = 1 ~ {\rm TeV} .
\end{eqnarray}

In order to obtain the correct DM abundance, see Eq.~(\ref{dmdensity}), one therefore needs to have:
\begin{eqnarray} \label{td2}
T_{\rm d} \gs 0.4-1.6 ~ {\rm GeV} ~ ~ ~ ~ ~ ~ m_\chi = 100 ~ {\rm GeV}-1 ~ {\rm TeV} \, .
\end{eqnarray}

For $T_{\rm d} \sim {\cal O}({\rm GeV})$, the corresponding modulus mass is found from Eq.~(\ref{Td}) to be
\be\label{mphi}
m_\phi \sim \mathcal{O}(1000) ~ {\rm TeV}\,,
\ee
with the exact value depending on the decay modes of the modulus.

Another non-thermal source of Higgsinos production is gravitino decay.
%For $m_{3/2} \gs 50$ TeV, gravitinos decay early enough not to affect the BBN. The main constraint on the abundance of heavy gravitinos comes from %the DM bound. Each gravitino decay eventually produces one DM particle.
Since gravitino decay occurs at a temperature $\ll {\cal O}({\rm GeV})$, and dark matter annihilation rate must satisfy the Fermi bounds~(\ref{fermi}), annihilation is very inefficient at this time. As a result, the density of Higgsinos produced from gravitino decay is therefore the same as that of the gravitinos. Therefore we must have:
\begin{eqnarray} \label{gravdens}
{n_{3/2} \over s} \ls 5 \times 10^{-10} ~ \left({1 ~ {\rm GeV} \over m_\chi}\right) \, .
%& \leq & 5 \times 10^{-12} ~ ~ ~ m_\chi \geq 100 ~ {\rm GeV} \, .
\end{eqnarray}

Gravitinos are produced via thermal and non-thermal processes in the early universe. Modulus decay dilutes gravitinos that were produced in the prior epochs (e.g., during inflationary reheating) by a huge factor. Thermal gravitino production after modulus decay is highly suppressed due to the low decay temperature $T_{\rm d} \sim {\cal O}({\rm GeV})$.

However, gravitinos can also be produced directly from modulus decay $\phi \rightarrow {\tilde G} {\tilde G}$.
%The decay to helicity $\pm 3/2$ states is helicity suppressed, which makes it negligible. The potential danger comes from helicity $\pm 1/2$ states, whose production can be enhanced at energies $m_\phi \gg m_{3/2}$, due to their goldstino nature, hence no helicity suppression.
%
The density of gravitinos thus produced
%Since two gravitinos are produced per modulus quanta in the $\phi \rightarrow {\tilde G} {\tilde G}$ process, we have
is given by $(n_{3/2}/s) = {\rm Br}_{3/2} (3 T_{\rm d}/4 m_\phi)$, where ${\rm Br}_{3/2}$ is the branching ratio for $\phi \rightarrow {\tilde G} {\tilde G}$ process. From Eqs.~(\ref{Td}), we then find
\begin{eqnarray} \label{gravbr}
{n_{3/2} \over s} \sim 5 \times 10^{-8} ~ \left(\frac{m_\phi}{100\, {\rm TeV}}\right)^{1/2} ~ {\rm Br}_{3/2} \, .
%{\rm Br}_{3/2} & \leq & 5 \times 10^{-5} ~ \left({100 ~ {\rm TeV} \over m_\phi}\right)^{1/2} \, \nonumber \\
%& \ls & {\cal O}(10^{-5}) ~ ~ ~ m_\phi \sim 2000-6000 ~ {\rm TeV} \, .
\end{eqnarray}

For the typical value of $m_\phi$ given in~(\ref{mphi}) and $100~{\rm GeV} \leq m_\chi \leq 1$ TeV, Eqs.~(\ref{gravdens},\ref{gravbr}) yield the following absolute upper bound:
\be \label{brconst}
{\rm Br}_{3/2} \ls 10^{-5} .
\ee
Any successful scenario for non-thermal Higgsino production from modulus decay must satisfy this limit.

\section{Non-thermal Higgsino Dark Matter: An Example} \label{KKLTexample}

%The problem can be overcome if $\phi$ does not dominate the energy density of the universe when it decays. In such a case, the right-hand side of Eqs.~(\ref{gravbr}) and~(\ref{brconst}) will be multiplied by $f_\phi$ and $f^{-1}_\phi$, respectively, where $f_\phi$ is the ratio of the energy density in $\phi$ to the total energy density of the universe at the time of decay. For $f_\phi < 10^{-3}$, the abundance of gravitinos will be suppressed to safe levels.

%To lower ${\rm Br}_{3/2}$ itself, one has to go beyond the KKLT model by considering scenarios with multiple moduli. Modifications to the modulus sector are needed to suppress the decay of the lightest modulus to helicity $\pm 1/2$ states of the gravitino. The condition for such a suppression are discussed in detail in~\cite{dine}.

As an explicit example, we take the case of mirage mediation in the context of KKLT compactification~\cite{Kachru:2003aw}. Working in $D=4,~N=1$ effective supergravity, the superpotential of the modulus sector consists of a flux term that fixes complex structure moduli, and a non-perturbative piece that fixes the K\"ahler modulus. The K\"ahler potential is given by $K = -3\ln{(T+\overline{T})} + (T+\overline{T})^{-n_m}\Phi \Phi^{\dagger}$, where $\Phi$ denotes matter fields and $n_m$ are the modular weights. The input parameters fixing the GUT scale masses are $m_{3/2}$, $\alpha$, $n_m$, and ${\rm tan}\beta$.

For our case study, we choose $n_m = 1/2$ for all matter fields and ${\rm tan}\beta = 50$. The general conclusions hold for other values of $n_m$ and ${\rm tan} \beta$.

\begin{figure}[!htp]
\centering
\includegraphics[width=3.5in]{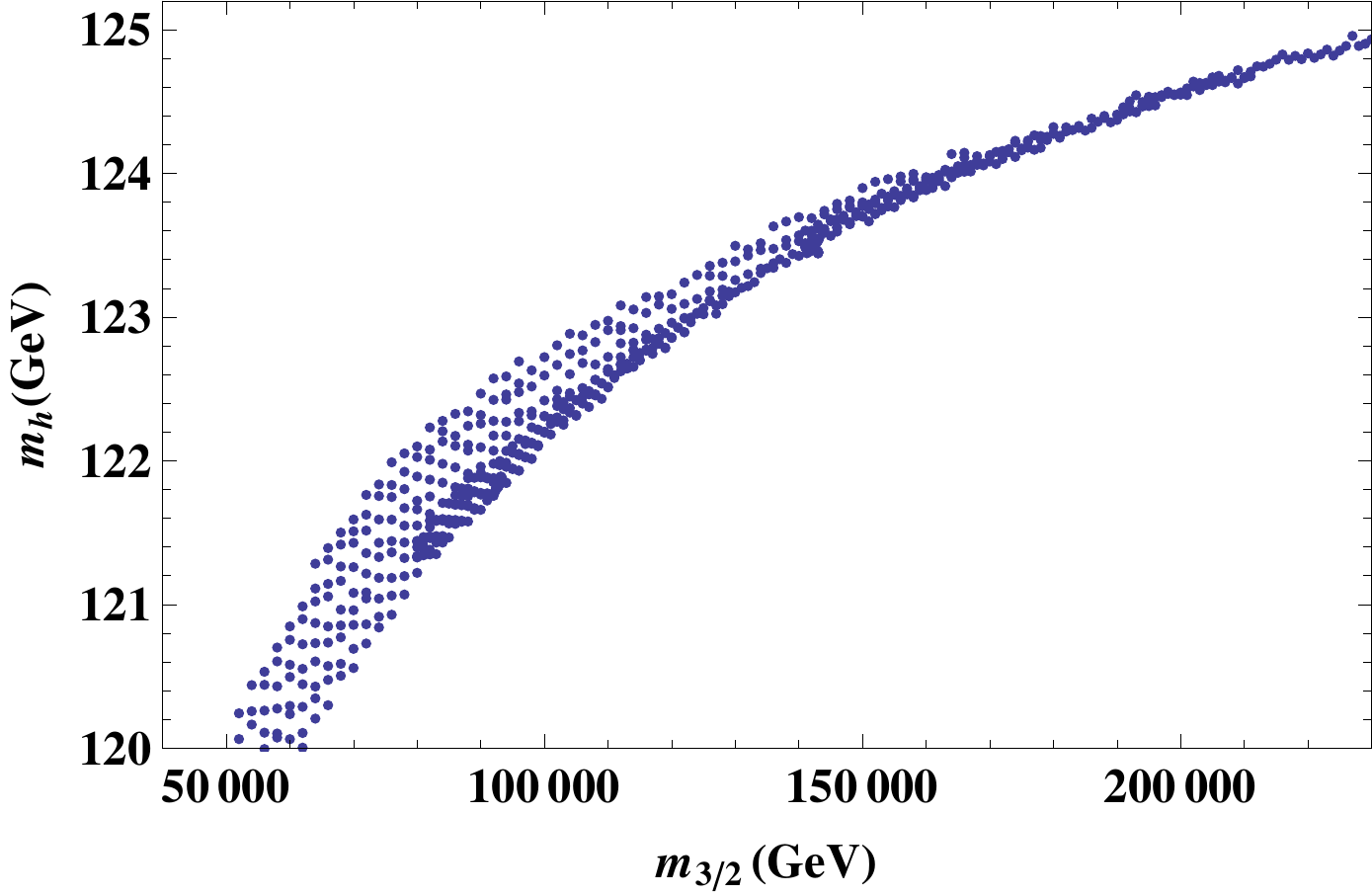}
\caption{Higgs mass versus gravitino mass. We choose $n_m =1/2$ and ${\rm tan}\beta = 50$, and scan over $\alpha$ and $m_{3/2}$. Heavy gravitinos decaying before the onset of BBN are typically compatible with a Higgs mass above 120 GeV. A similar behavior is obtained for other values of ${\rm tan}\beta$ and $n_m$}
\label{higgsmassgravitinomass}
\end{figure}

The scalar spectrum has the suppression given in Eq.~(\ref{hierarchy}) with respect to the gravitino mass. It is instructive to note that when the stops are themselves hierarchically related to the gravitino as in this case, $m_h \sim 125$ GeV is compatible with heavy gravitinos that decay before the onset of BBN. Since the one-loop correction to the Higgs mass depends logarithmically on $m_{3/2}$, a little heavier Higgs is preferred by the cosmologically safe region. We plot the dependence of the Higgs mass on the gravitino mass in Figure \ref{higgsmassgravitinomass}.

\begin{figure}[!htp]
\centering
\includegraphics[width=3.5in]{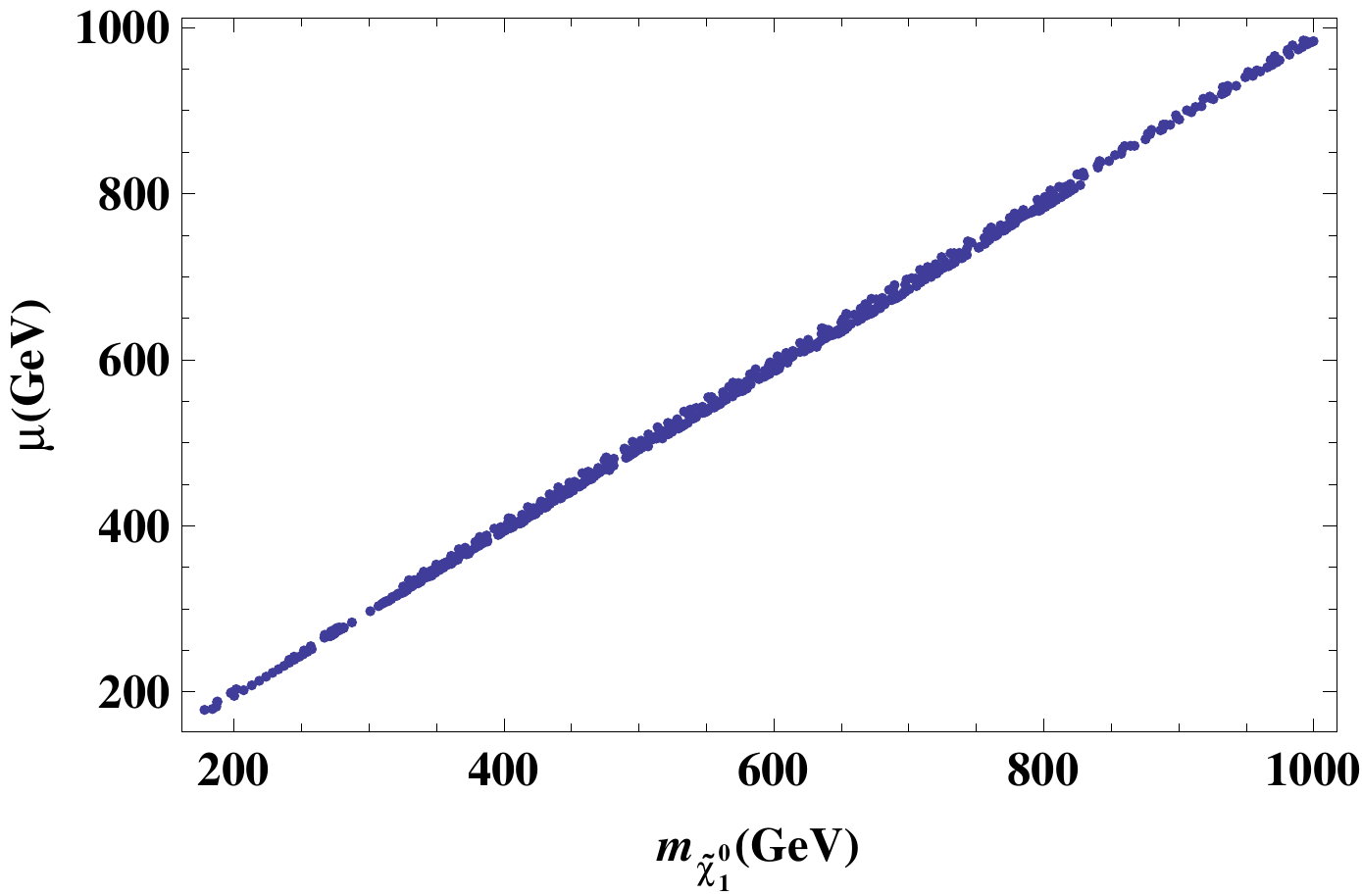}
\caption{$\mu$ versus LSP mass. For sub-TeV LSP, the dark matter is always a Higgsino.}
\label{massneutralinomu}
\end{figure}

For given $n_m =1/2$ and ${\rm tan}\beta = 50$, scanning over $0.1 < \alpha < 1.6$, and $m_{3/2} > 40$ TeV, one finds that for LSP mass below $\sim 1$ TeV, the LSP is always a Higgsino. We plot $\mu$ against LSP mass in Figure \ref{massneutralinomu}. This result is independent of the choice of ${\rm tan}\beta$. Since the gaugino mass does not depend upon $n_m$, this conclusion is also independent of $n_m$.

\begin{figure}[!htp]
\centering
\includegraphics[width=3.5in]{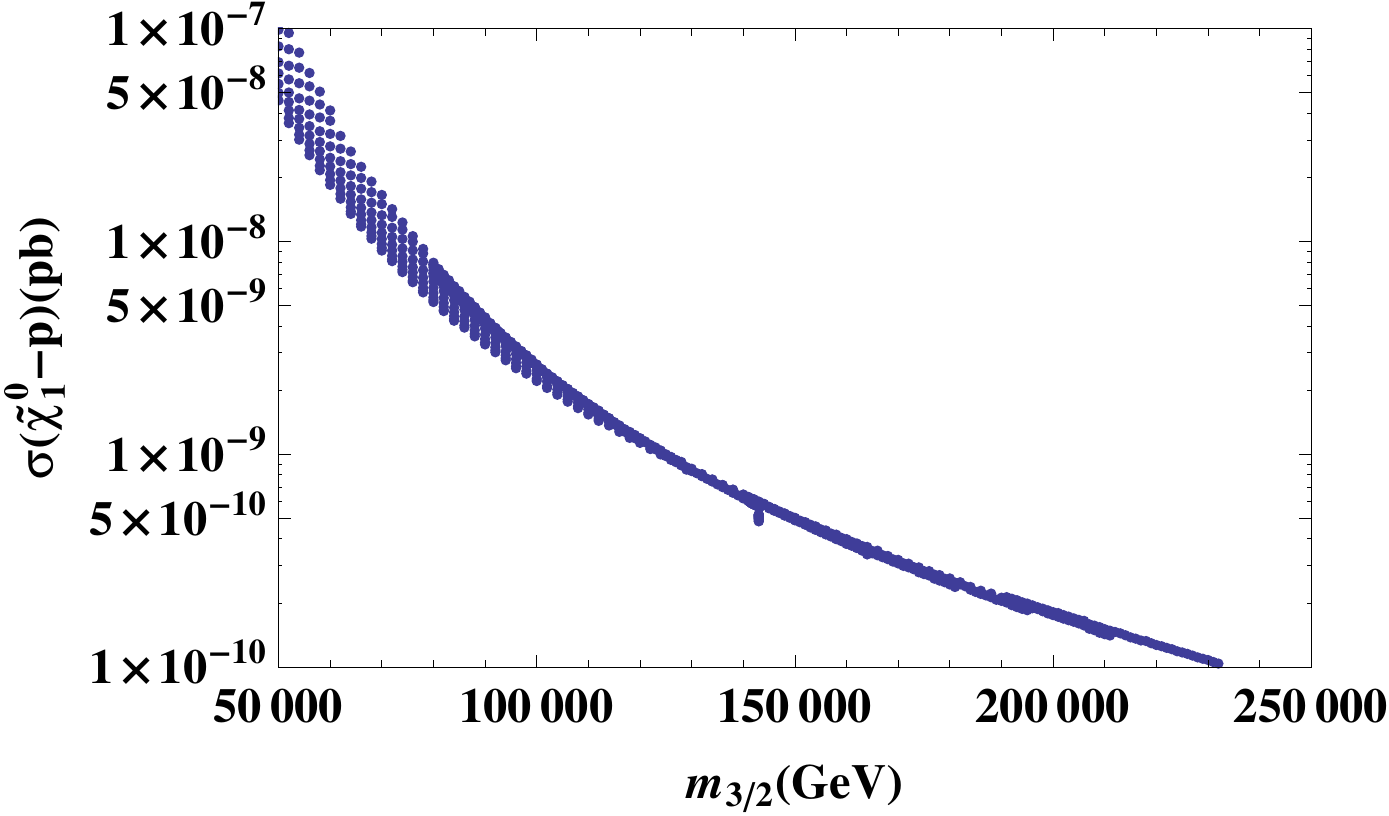}
\caption{Spin independent scattering cross section versus gravitino mass. For values of $m_{3/2}$ that are compatible with 125 GeV Higgs, see Fig. 1, $\sigma_{{\tilde \chi}^1_0-p}$ satisfies the experimental data.}
\label{protonneutralinocrosssectiongravitino}
\end{figure}

We also plot the spin independent scattering cross section $\sigma_{{\tilde \chi}^0_1-p}$ for various values of the gravitino mass in Figure \ref{protonneutralinocrosssectiongravitino}. Since larger gravitino mass is correlated to a larger heavy Higgs ($H$) mass in this model, $\sigma_{\tilde \chi^0_1-p}$ becomes smaller as $m_{3/2}$ increases. Again, this is compatible with cosmologically safe region and $m_h \sim 125$ GeV. The current bound on the cross section is $\sim 2 \times 10^{-9}$ pb for a dark matter mass of $55$ GeV~\cite{direct}, and relaxes as dark matter mass increases.

%The gravitino, gaugino, and scalar spectra have the correct hierarchies and the modulus mass is in the correct range.

%First, we demonstrate the logarithmic dependence of the mass of the Higgs on the gravitino mass below. For simplicity, we take $\kappa_0 = \kappa_A$, by taking the relevant modular weights to be equal.

%\textbf{*********Show gravitino-Higgs plot here******}.

%Clearly, a $125$ GeV Higgs prefers gravitino mass $\geq 40$ TeV.

%In the simplest cases of KKLT compactification or LVS compactification the mass of the modulus is given by
%
%\be
%m_{\phi} \, \sim \, \mathcal{O}(1-2)\ln (m_{3/2}) \times m_{3/2}\,\,.
%\ee
%

%We plot the enhancement factor $T_f/T_r$ as a function of the Higgs mass below.

Table \ref{KKLTbenchmark} depicts a few benchmark points of consistent non-thermal Higgsino dark matter in this model. In the table, we show the annihilation rates at present $\langle \sigma_{\rm ann} v \rangle_0$ and at the time of freeze-out $\langle \sigma_{\rm ann} v \rangle_f$. We see that for all of these points $\langle \sigma_{\rm ann} v \rangle_0$ satisfies the Fermi bounds. 

Dark matter annihilation at the freeze-out occurs mostly through $S$-channel and a coannhilation component. The latter arises due to the fact that masses of the second lightest neutralino and chargino are close to LSP mass $m_{{\tilde \chi}^1_0}$.
The dark matter content of the universe is obtained by multiplying the $\langle \sigma_{\rm ann} v \rangle_f$ by $T_{\rm f}/T_{\rm d}$.
We have taken $c=0.4$ to calculate $T_{\rm d}$, which is the leading order value appearing in the decay width of the modulus in this particular example, in the limit of $\alpha = 1$ corresponding to zero dilaton-modulus mixing in the gauge kinetic function. But $c$ can also be $\sim 1$ depending on relative contributions of the modulus and dilaton in the gauge kinetic function, and $T_{\rm d}$ can be $\mathcal{O}(1-2)$ of its central value. With this taken into account, it is seen that $T_{\rm f}/T_{\rm d}$ is in the right range to yield the correct dark matter content of the universe.
%The DM content of the universe is obtained by multiplying $\langle \sigma_{\rm ann} v \rangle_f$ by $T_{\rm d}/T_{\rm f}$.
%Dark matter annihilates mostly through $S$-channel and coannhilation. The coannihilation component arises due to the fact that masses of the second %lightest neutralino and chargino are close to LSP mass. In the table we show the annihilation cross sections at present and at the time of %freeze-out.

We also show the value of $\sigma_{\tilde\chi^0_1-p}$ for these points, which are well allowed by the experimental data.
The table also includes masses of the gluino and stops, and the Higgs mass $m_h$ for these points.

%\textbf{********* Show Higgs 124.5 point here***********}
%\begin{table}[!htp]
%\caption{Benchmark scenarios with non-thermal Higgsino dark matter. All masses are in GeV.}
%\label{KKLTbenchmark}
%\begin{center}
%\begin{tabular}{c c c c c} \hline \hline
%  Higgs           & $\tilde{\chi}^0_1$       & $T_f/T_r$        & Current $\sigma_{\rm ann}$  &  $\sigma_{\tilde{\chi}^0_1-p}$           \\ \hline \hline
%  $123.5$         &                          &                  &                             &                          \\
%  $124.5$         &                          &                  &                             &   \\
%  $125.5$         &                          &                  &                             &   \\ \hline \hline
%\end{tabular}
%\end{center}
%\end{table}

%\textbf{*******Show neutralino-proton cross-section vs. Higgs plot here ********}.
%\textbf{*******Other vs. Higgs plots here ********}.

\section{General Conditions for suppressing gravitino production} \label{gravitino}

%Condition $1$: \textit{}

%The first general condition is that anomaly mediation contributions should be competitive with

In the KKLT model discussed above, the partial decay rate for $\phi \rightarrow {\tilde G} {\tilde G}$ is $\Gamma_{3/2} = m^3_\phi/288 \pi M^2_{\rm P}$. Then, after using Eq.~(\ref{decaywidth}), we find that ${\rm Br}_{3/2} \sim {\cal O}(10^{-2})$. This implies that gravitino decay will overproduce Higgsinos by 3 orders of magnitude in this model, see Eq.~(\ref{gravbr}). The main reason for obtaining such a large ${\rm Br}_{3/2}$ is that modulus decay to helicity $\pm 1/2$ gravitinos is not helicity suppressed in the KKLT model \cite{Dine}.

The problem can be overcome if the modulus $\phi$ does not dominate the energy density of the universe when it decays. In such a case, the right-hand side of Eqs.~(\ref{gravbr}) and~(\ref{brconst}) will be multiplied by $f_\phi$ and $f^{-1}_\phi$, respectively, where $f_\phi$ is the ratio of the energy density in $\phi$ to the total energy density of the universe at the time of decay. For $f_\phi < 10^{-3}$, the abundance of gravitinos will be suppressed to safe levels.

Alternatively, one can seek conditions for suppressing gravitino production from modulus decay (for example, see \cite{Dine}). Here we briefly outline general conditions for such a suppression, and stress that the main ingredients of a successful scenario for non-thermal Higgsino production presented above should also hold in cases where the gravitino production is suppressed.

\onecolumngrid

\begin{table}[!htp]
\caption{Some benchmark points of non-thermal Higgsino dark matter for mirage mediation model in the context of KKLT compactification. The input parameters are $\alpha$ and $m_{3/2}$. The modular weights are fixed to be $n_m = 1/2$, and ${\rm tan}\beta = 50$.
%The spin independent scattering cross section is $\sigma_{\tilde{\chi}^0_1-p}$.
All masses are in GeV.}
\label{KKLTbenchmark}
\begin{center}
\begin{tabular}{c c | c c c c c c c c c c} \hline \hline

%\, \,  & \, \,  & \, \,  & \, \,  & \, \, & \, \,  & \, \,  & \, \,  & \, \, & \, \, & \, \, & \, \,   \\

$\alpha$ \,\, & $m_{3/2}$ \,\, & $m_h$ \,\, & $m_{\tilde{\chi}^0_1}$ \,\, & $m_{\tilde{\chi}^0_2}$\,\, & $m_{\tilde{g}}$ \,\,& $m_{\tilde{t}_1}$ \,\,& $m_{\tilde{t}_2}$ \,\,  & $\langle \sigma_{\rm ann} v \rangle_{0}$ ({$\rm cm^3/s$})\,\, & $\langle \sigma_{\rm ann} v \rangle_{f}$ ({$\rm cm^3/s$}) \,\, & $T_{\rm f}/T_{\rm d}$ \,\,       &  $\sigma_{\tilde{\chi}^0_1-p}$ (pb)           \\ \hline \hline
% \, \, & \, \, & \, \, & \, \, & \, \, & \, \, & \, \, & \, \, & \, \, & \, \, & \, \, & \, \, \\
$1.49$ \,\, & $143 \cdot 10^3$ \,\, & $123.5$ \,\, & $248.4$ \,\,& $250.8$ \,\, & $3828$ \,\,& $2441$ \,\,& $2781$ \,\, & $1.49 \cdot 10^{-25}$ \,\,& $1.63 \cdot 10^{-25}$ \,\, & $\sim 6$ \,\,& $5 \cdot 10^{-10}$ \\

$1.46$ \,\, & $200 \cdot 10^3$ \,\, & $124.5$ \,\, & $258.9$ \,\,& $260.6$ \,\, & $5536$ \,\,& $3564$ \,\, &$3991$ \,\,& $1.38 \cdot 10^{-25}$ \,\, & $1.52 \cdot 10^{-25}$ \,\,& $\sim 3.4$ \,\,& $1.4 \cdot 10^{-10}$  \\

$1.44$ \,\, & $232 \cdot 10^3$ \,\, & $125$ \,\, & $306$ \,\,& $308$ \,\, & $6505$ \,\,& $4197$ \,\, &$4677$ \,\,& $1.01 \cdot 10^{-25}$ \,\, & $1.01 \cdot 10^{-25}$ \,\,& $\sim 3.2$ \,\,& $8.9 \cdot 10^{-11}$  \\   \hline \hline

%  $123.5$         & $248$                    &                  &                             &                          \\
%  $124.5$         &                          &                  &                             &   \\
%  $125.5$         &                          &                  &                             &   \\ \hline \hline
\end{tabular}
\end{center}
\end{table}

\twocolumngrid

The decay of a modulus to other fields depends on the interaction terms in the Lagrangian, and the requirement for suppressing decay to gravitinos will be reduced to a set of constraints in the effective theory. To have a more concrete demonstration of what kinds of constraints may emerge, we choose to work in effective supergravity, with a modulus coupling to the visible sector through the gauge kinetic function. This is the scenario in the class of Type IIB models discussed above.

In general, one can consider a scenario with multiple moduli $\phi_i$, with the decaying modulus appearing in the gauge kinetic function. The normalized eigenstates $\phi_n$ are given by
\be \label{norm}
(\phi)_i \,\, = \,\, \sum_j \,\, C_{ij} \,(\tau_n)_j \,\,\,,
\ee
where the $C_{ij}$ are eigenvectors of the matrix $K^{-1} \,\partial^2 V$. For simplicity, we will assume diagonal $C_{ij}$ with entries $C_i$. The partial widths for modulus decay to gauge fields, gauginos and helicity $\pm 1/2$ gravitinos are
%
%
%\begin{eqnarray}
%\mathcal{L} &=& \frac{1}{4} \epsilon^{k\ell mn} \left( G_{,T_i} \partial_k T - G_{,T^*_i} \partial_k T^{*} \right)
%\bar\psi_\ell \bar\sigma_m \psi_n \nonumber \\
%&-& \frac{1}{2} e^{G/2} \left( G_{,T_i} T + G_{,T^*_i} T^{*}_i \right) \left[ \psi_m \sigma^{mn} \psi_n + \bar\psi_m \bar\sigma^{mn} \bar\psi_n \right], \nonumber\\
%\end{eqnarray}
%
%where $G =  K +\log |W|^2$ is the K\"ahler function. The decay width to helicity $\pm 1/2$ components is given by
%
\bea \label{GammaTGravitino}
\Gamma_{\phi_i \rightarrow g g} & = &  \frac{N_g}{128\pi} ~ \frac{1}{\Expect{\tau}^{2}} \, C_i^2 \frac{m_{\phi_i}^3}{M^2_{\rm P}} \nonumber \\
\Gamma_{\phi_i \rightarrow \tilde{g} \tilde{g}} \, & = & \, \sum_p \frac{N_g}{128\pi} ~ C_p^2 ~ \Expect{\partial_p F^{i}}^2 \frac{m_{\phi_i}}{M^2_{\rm P}} \nonumber \\
\Gamma_{\phi_i \rightarrow {\tilde G} {\tilde G}} & \sim & \frac{1}{288\pi} ~ \left(|G_{\phi_i}|^2 K_{\phi_i\bar{\phi_i}}^{-1}\right) \frac{m^2_\phi}{m_{3/2}^2}\frac{m_{\phi_i}^3}{M^2_{\rm P}} \,\, . \nonumber \\
\,
\eea
where $G =  K +\log |W|^2$ is the K\"ahler function.

Under suitable choices of the K\"ahler potential, the required condition ${\rm Br}_{3/2} \sim 10^{-5}$ may be obtained. Similarly, the decay temperature of the modulus may be obtained in terms of the K\"ahler potential and superpotential. We refer to \cite{Allahverdi:2010rh} for more details.

For a single modulus, the branching ratios to gauge bosons and gauginos are roughly equal, and the branching to the gravitino needs to be suppressed, leading to the condition
\be
\frac{m_{\phi}}{m_{3/2}}|G_{\phi}| K_{\phi \bar{\phi}}^{-1/2} \, \sim \, 10^{-3} \,\,.
\ee

For the KKLT example, the above quantity is $\mathcal{O}(1)$, which leads to overproduction of gravitinos. However, in a more general scenario, one can suppress this ratio to the required levels by a suitable choice of $K_{\phi \bar{\phi}}$ and vacuum expectation value of $\phi$. This does not necessarily affect the existence of other conditions for successful non-thermal Higgsino production, such as comparable anomaly mediated contributions, or a modulus in the correct mass range. Moreover, the scalar masses depend on the holomorphic bisectional curvature of the plane (in tangent space) spanned by the scalars and the supersymmetry breaking modulus, and this is not necessarily changed by a shift in the metric $K_{\phi \bar{\phi}}$.
One can therefore expect to have a viable non-thermal scenario with the Higgs mass $m_h \sim 125$ GeV, while suppressing gravitino production from modulus decay.
%Thus, the stops can be in the range acceptable for a $125$ GeV Higgs, in a non-thermal Higgsino dark matter scenario.
We leave the detailed exploration of these issues for future work.

%Higgs-motivated approach to non-thermal Higgsino dark matter may still be valid in general.
%For a single modulus with $K = -3 \ln{\phi}$, one obtains ${\rm Br}_{3/2} \sim 10^{-2}$.

\section{Conclusions}

Considering dark matter in the sub-TeV range, thermal freeze-out underproduces Higgsino-like LSP. It is possible to enhance the relic density using non-thermal mechanisms of dark matter production. The enhancement, however, needs to obey the constraints from the gamma ray flux from dwarf spheroidal galaxies and the dark matter content of the universe. Moreover, there should be no overproduction of dark matter at any later stage, for example by the decay of gravitinos.

In this paper, we demonstrated these ideas in a generic class of mdoels where anomaly and modulus mediated contributions to supersymmetry breaking are comparable. Interestingly, we found that within this class of models, heavy gravitinos that are not subject to BBN bounds force the Higgsino as the only viable dark matter candidate in the sub-TeV range. We considered an explicit example of mirage mediation model in $D=4,~N=1$ supergravitiy arising from type IIB KKLT compactification, where the modulus decay provides the non-thermal origin of Higgsino-like dark matter. The large gravitino mass is helpful to yield $m_h$ around $125$ GeV in this model and satisfy the constraints arising from the dark matter direct detection experiments. We also discussed the general conditions to avoid the overproduction of LSP from gravitino decay in such scenarios.

%%%%%%%%%%%%%%%%%%%%%%%%%%%%%%%%%%%
\section{Acknowledgement}

This  work is partially  supported in part by the DOE grant DE-FG02-95ER40917. We thank Sheldon Campbell for very useful discussions. We thank the Center for Theoretical Underground Physics and Related Areas (CETUP* 2012) in South Dakota for its hospitality and for partial support during the completion of this work.

%%%%%%%%%%%%%%%%%%%%%%%%%%%%%%%%

\end{document}